\documentclass[dvips,prl,tightenlines,showpacs,twocolumn]{revtex4}

\usepackage{graphicx}
\usepackage{bbm}
\usepackage{amsfonts}
\usepackage{amsthm}
\usepackage{amssymb}
\usepackage{amsmath}

\newcommand{\id}{\mathbbm{1}}
\newcommand{\W}{{\mathcal W}}
\newcommand{\beq}{\begin{eqnarray}}
\newcommand{\eeq}{\end{eqnarray}}
\newcommand{\ist}{\;=\;}

\begin{document}
\title{Two distinct classes of bound entanglement: PPT--bound and ``multi-particle''--bound}
\author{Beatrix C. Hiesmayr and Marcus Huber\\
    \small{Faculty of Physics, University of Vienna,
    Boltzmanngasse 5, A-1090 Vienna, Austria}}

\begin{abstract}
We introduce systematically with the help of Weyl operators novel
classes of multipartite and multidimensional states which are all
bound entangled for arbitrary dimension. We find that the
entanglement is bound due to different reasons: unlockable due to
the multi--particle nature and some states are in addition bound due
the fact being positive under partial transposition (PPT). By a
general construction ($\W$ simplices) we obtain classes of states
which have the same geometry concerning separability and
entanglement independent of the number of involved particle pairs.
Moreover, we introduce a distillation protocol and demonstrate for
$d=3$ that for a certain set of states the entanglement can be
increased only up to a certain amount. \pacs{03.67.Mn}
\end{abstract}
\maketitle


Quantum entanglement is a key feature of quantum theory with many
important consequences for modern physics. It has become a highly
valuable resource for novel applications, such as cryptography and a
formidable quantum computer. However, the mathematical and/or
physical characterization of all types of entanglement and their
implementations are far from being fully explored. E.g. the
quantification or even the classification of entanglement of
multipartite systems is still an open problem.

This paper will analyze the nature of at least two distinct classes
of bound entanglement, i.e. entanglement which cannot be distilled
by local operations and classical communication (LOCC) into pure
maximally entangled states, when each local observer posses only one
particle. This in return means that there should exists different
applications for these states due to the different nature of their
entanglement.

We first review a huge class of bipartite qudit states. A qudit is a
quantum systems with $d$ degrees of freedom. With the help of group
theoretical methods which allows for considerable simplifications a
geometrical picture of the state space can be drawn, i.e. the
properties separability, \textit{bound} entanglement or PPT
entanglement (PPT$\ist$positive under partial transposition) and NPT
entanglement (NPT$\ist$negativ under partial transposition) can be
characterized. For bipartite qudits this state space was called
``magic'' simplex $\W$ in Ref.~\cite{BHN1} and extensively discussed
in Refs.~\cite{BHN2,BHN3} in different contexts. The construction of
a simplex of states with maximally mixed subsystems has so far
proven to be a powerful tool in analyzing bipartite qubits and
qutrits (e.g. Ref.~\cite{Derkacz,Krammer1,Krammer2}) and recently
even for multipartite qubits~\cite{HHHKS1}. It provides a deep
insight into the structure of entangled states and helps in
constructing entanglement witnesses and exploring entanglement
measures.

We will extend the simplex of bipartite qudits, i.e. one pair of
qudits, to $n$ pairs of qudits where $n$ is any natural number. We
will prove that interestingly this extended class has the same
properties concerning separability, bound entanglement and
NPT--entanglement by proving that the optimal entanglement witnesses
reduces to the same mathematical conditions (Theorem $2$).
Therefore, results for bipartite qudits become automatically true
for any $n$ pairs of bipartite qudits, which may otherwise due to
the high computational effort would not be obtainable.

This extended class of states shows due to their multi-particle
nature a feature which was called unlockable--bound entanglement
\cite{Smolin,HorodeckigeneralizedSmolin2}. In detail
$NPT$--entangled states can be distilled to certain extremal states,
the so called ``vertex'' states of the simplex $\W^{\otimes n}$,
however, not into pure maximally entangled states: this novel class
of states are bound to their own class. For multipartite qubits this
was shown in Refs.~\cite{HHHKS1}. We prove in this paper that this
is a general feature of such multipartite simplex states and,
moreover, the fact that $PPT$--bound entangled states exist for
dimensions $d\geq 3$ implies that there are two different kinds of
bound entanglement. Explicitly, we give a multidimensional
distillation protocol for $d=3$ which distills certain states within
the simplex to the vertex states, which are themselves bound
entangled.


\begin{figure*}
\begin{center}
(a)\includegraphics[width=4cm, keepaspectratio=true]{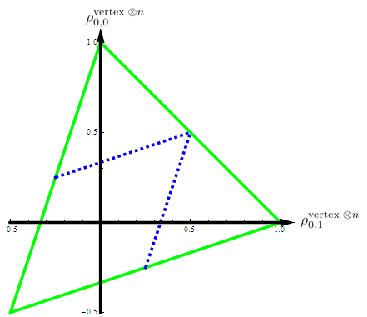}
(b)\includegraphics[width=4.4cm, keepaspectratio=true]{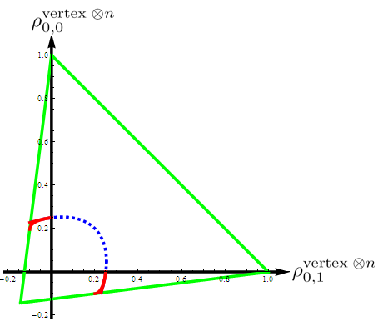}
(c)\includegraphics[width=4cm, keepaspectratio=true]{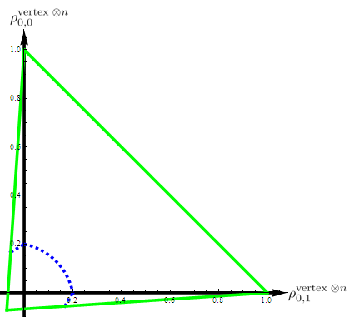}
\caption{(Color online) Slices via the simplices for the states
which are mixtures of any two vertex states and the maximally mixed
state, Eq.~(\ref{exampledim}), for the dimensions (a) $d=2$, (b)
$d=3$, (c) $d=4$. The (green) triangles are given by the positivity
condition, the dotted (blue) lines/curves represents the PPT
condition. For $d=3$ one finds a whole region of PPT bound
entanglement if either $\alpha$ or $\beta$ is negative (filled (red)
region). As expected the region of separable states shrinks with
increasing dimension $d$.}\label{linestates}
\end{center}
\end{figure*}

\textbf{The magic simplex $\W$ for bipartite qudits:}
For bipartite qudits the vertex states $P_{i,j}$ of the ``magic''
simplex $\W$ are the maximally entangled states in $d$ dimensions
(Refs.~\cite{BHN1,BHN2}):
\begin{eqnarray}
|\Phi^+\rangle:&=&\sum_{i=0}^{d-1}|ii\rangle\;,\quad
P_{0,0}\;:=\;|\Phi^+\rangle\langle\Phi^+|\\
P_{k,l}:&=&\id_d\otimes {\mathrm W_{k,l}}\;P_{0,0}\;\id_d\otimes
{\mathrm W}^\dagger_{k,l}
\end{eqnarray}
where the $\mathrm W_{k,l}$ are the Weyl operators defined by
\begin{eqnarray}
{\mathrm W}_{k,l}|s\rangle\;=\;
w^{k(s-l)}\;|s-l\rangle\qquad\textrm{with}\quad w\;=\;e^{2\pi
i/d}\;.
\end{eqnarray}
The magic simplex $\W$ is the convex combination of all vertex
states
\begin{eqnarray}
\W:=\left\lbrace\sum_{k,l=0}^{d-1} c_{k,l}\; P_{k,l}\;|\;c_{k,l}\geq
0,\;\sum_{k,l=0}^{d-1} c_{k,l}=1\right\rbrace\;.
\end{eqnarray}
One main property of this class of states forming a $d^2-1$
dimensional simplex is that any trace of one particle results in a
maximally mixed state. We want to conserve this property for the
multipartite scenario, i.e. any trace of one or more particles
should result into a maximally mixed state:
\beq\label{generalizedsmolin}\rho^{\textrm{vertex}\otimes
n}_{0,0}&:=&\frac{1}{d^2}\sum_{i,j=0}^{d-1} P_{i,j}\otimes
P_{i,j}\otimes\dots\otimes
P_{i,j}\nonumber\\&\ist&\frac{1}{d^2}\sum_{i,j=0}^{d-1}
P_{i,j}^{\otimes n}\;.\eeq For $d=2$ this state was investigated by
Smolin~\cite{Smolin} and has proven to be an interesting state,
exhibiting many counter-intuitive properties: such that it is
biseparable under any bipartite cut, but ignorance of any arbitrary
number of subsystems will render this state useless for quantum
informational tasks,
Refs.~\cite{HorodeckigeneralizedSmolin,HorodeckigeneralizedSmolin2},
though it violates a Bell inequality (see
Refs.~\cite{Dur,HHHKS1,HorodeckigeneralizedSmolin}). Moreover,
applying the two sets of multipartite entanglement measures proposed
in Ref.~\cite{HHK1} it turns out that $n$ paired LOCCs are needed to
prepare the state, whereas $2n$ parties are needed to cooperate
locally to perform quantum informational tasks with that state.



We prove now that the above state has unlockable entanglement and
then generalize to a whole class of states with all that features.\\
\\
\textbf{Theorem 1:} The state, Eq.~(\ref{generalizedsmolin}), is
(multipartite) bound entangled for any dimension $d$, because no
locally working of all involved parties can by LOCC distill a pure
maximally entangled state.
\begin{proof}
As for every state that exhibits a partial separability like
$
\rho=\sum_i p_i\; \rho^A_i\otimes\rho^B_i
$
will remain $A$-$B$ separable under every LOCC of the form:
\begin{eqnarray}
\Lambda_{LOCC}[\rho]=\frac{\sum_k A_k\otimes B_k\;\rho\; A^\dagger_k\otimes B^\dagger_k}{\text{Tr}(\sum_k A_k\otimes B_k\;\rho\; A^\dagger_k\otimes B^\dagger_k)}
\end{eqnarray}
and the state in question allows a biseparable decomposition even if
two subsystems are arbitrarily exchanged, this special property is
preserved under LOCC. No maximally entangled pure state can exhibit
this property, hence the state is bound entangled.
\end{proof}

\textbf{The magic simplex $\W^{\otimes n}$ for $n$ pairs of qudits:}
A certain vertex states of any $n$ pairs of qudits can be defined by
\beq n=1:\qquad\rho^{\textrm{vertex}\otimes
1}_{0,0}&:=&P_{0,0}\nonumber\\
n\geq 2:\qquad\rho^{\textrm{vertex}\otimes
n}_{0,0}&:=&
\frac{1}{d^2}\sum_{i,j=0}^{d-1} P_{i,j}^{\otimes n}\;.\eeq By
applying in one subsystem a Weyl operator $
W_{k,l}:=\id_{d}\otimes{\mathrm W}_{k,l}$ one obtains as before the
remaining $d^2-1$ vertex states \beq \rho^{\textrm{vertex}\otimes
n}_{k,l}&=& 
\id_{d^2}^{\otimes (n-1)}\otimes W_{k,l}\;
\rho^{\textrm{vertex}\;\otimes n}_{0,0}\;\id_{d^2}^{\otimes
(n-1)}\otimes W_{k,l}^\dagger\nonumber\\
&=&\frac{1}{d^2}\sum_{i,j=0}^{d-1} P_{i,j}^{\otimes n-1}\otimes
W_{k,l} P_{i,j} W_{k,l}^\dagger\;.\eeq Note that if the Weyl
operator is applied on a different subsystem we obtain an equivalent
simplex, however, with different labeling (all states and partial
states have for any $n$ same eigenvalues).

Now we can define a huge class of states which have the same
geometry concerning separability and entanglement for a given $d$,
the ``magic'' $n$ pair qudit simplex $\W^{\otimes n}$:
\begin{eqnarray}
\W^{\otimes n}:=\left\lbrace\sum_{k,l=0}^{d-1} c_{k,l}\;
\rho^{\textrm{vertex}\;\otimes n}_{k,l}\;|\;c_{k,l}\geq
0,\;\sum_{k,l=0}^{d-1} c_{k,l}=1\right\rbrace\;.\nonumber
\end{eqnarray}

These states have the same properties as the vertex states, i.e. all
subsystems are maximally mixed, all states have $n$--separable
decompositions, where always any two subsystems can be grouped
together and single subsystems may arbitrarily be interchanged. The
mixedness of any vertex state,
$M:=\frac{d}{d-1}(1-\text{Tr}(\rho^{\textrm{vertex}\;\otimes
n}_{k,l}\rho^{\textrm{vertex}\;\otimes n}_{k,l}))$,  for $n\geq2$ is
$\frac{1-d^{-2}}{1-d^{-n}}$, thus gets less mixed with increasing
$n$ and/or $d$.

We prove now that the structure of separability is for any $n$
equivalent by the powerful tool of witnesses, then we proceed to
discuss the feature of bound entanglement and unlockable--bound
entanglement.

\textbf{Optimal witnesses in the simplex $\W^{\otimes n}$:} An
entanglement witness $EW_\rho$ is a criterion to ``witness'' for an
certain state $\rho$ that it is not in the set of separable states
$SEP$. 
Knowing that SEP is convex it can be
completely characterized by the tangential hyperplanes, thus we
search for tangential or optimal witnesses on the surface of SEP,
i.e.
\begin{eqnarray}
EW^{opt}_\rho\ist\lbrace  K\;=\; K^\dagger\;\not=\;
0|\forall\;\rho_{sep}\in SEP:\nonumber\\  Tr(K\;\rho_{sep})\;<\;
0\quad\textrm{and}\quad Tr(K\;\rho)\ist 0\rbrace\;.
\end{eqnarray}
As proven in Ref.~\cite{BHN1} any witness operator for states within
the simplex $\W$ can only be of the form $
K\ist\sum_{k,l}\kappa_{k,l}\; P_{k,l}$. As $\W$ and $\W^{\otimes n}$
have the same group symmetries by their construction via the Weyl
operators (see Theorem $6$ in Ref.~\cite{BHN1}) any witness operator
within $\W^{\otimes n}$ has to have the form $K_n\ist\sum_{k,l}
\kappa_{k,l} \rho^{\textrm{vertex}\;\otimes n}_{k,l}$.\\
\\
\textbf{Theorem 2:} The operator $K_n=\sum_{k,l} \kappa_{k,l}
\rho^{\textrm{vertex}\;\otimes n}_{k,l}$ is an optimal entanglement
witness if $\det{M_\Phi}\ist 0$ with $M_\Phi\ist \sum_{k,l}
\kappa_{k,l} W_{k,l}|\Phi\rangle\langle\Phi|W_{k,l}^\dagger\geq
0\qquad\forall\quad\Phi\in\mathbb{C}^d\;.$ This means that the set
of separable, PPT--entangled and NPT--entangled states have for any
$d$ and all $n$ the same geometry because the $d\times d$ matrix
$M_\Phi$ is identical.

\begin{proof}
Any separable state $\rho_{sep}$ can be written as a convex
combination of pure product states and therefore
$Tr(K_n\;\rho_{sep})\geq 0\quad \forall\quad \rho_{sep}\in SEP$
implies that
\begin{widetext}
\beq \langle
K_n\rangle\;:=\;\langle\eta_1,\chi_1|\otimes\langle\eta_2,\chi_2|\otimes\dots\langle\eta_n,\chi_n|\;K_n\;|\eta_1,\chi_1\rangle\otimes
|\eta_2,\chi_2\rangle\otimes\dots|\eta_n,\chi_n\rangle\geq
0\qquad\forall\quad\eta_1,\chi_1,\eta_2,\chi_2\cdots\eta_n,\chi_n,\in\mathbb{C}^d\;.\nonumber\eeq
By the observation that $P_{k,l}\ist\frac{1}{d}\sum_{s,t=0}^{d-1}
W_{k,l}\;|ss\rangle\langle
t,t|\;W_{k,l}^\dagger\ist\frac{1}{d}\sum_{s,t}
\mathrm{W}_{k,l}\otimes\id_d\;|ss\rangle\langle
t,t|\;\mathrm{W}_{k,l}^\dagger\otimes\id_d$ follows \beq
\langle\eta_i,\chi_i|\;P_{k,l}\;|\eta_i,\chi_i\rangle&=&\frac{1}{d}\sum_{s,t}\langle\eta_i|\mathrm{W}_{k,l}|s\rangle
\langle\chi_i|s\rangle\langle t|\chi_i\rangle\langle
t|\mathrm{W}_{k,l}^\dagger|\eta_i\rangle
\ist\frac{1}{d}\langle\eta_i|\mathrm{W}_{k,l}|
\phi_i\rangle\langle\phi_i|\mathrm{W}_{k,l}^\dagger|\eta_i\rangle\;,
\eeq where we defined all $\phi_i\in\mathbb{C}^d$ as
$|\phi_i\rangle\ist\sum_{s}\langle\chi_i|s\rangle\,|s\rangle$.
Therefore, $P_{k,l}$ is obviously an entanglement witness,
because\beq
\frac{1}{d}\langle\eta_i|\mathrm{W}_{k,l}|\phi_i\rangle\langle\phi_i|\mathrm{W}_{k,l}^\dagger|\eta_i\rangle\;
=\frac{1}{d}|\langle\eta_i|\tilde\phi_i\rangle|^2\geq\;0\;,\forall\;\eta_i,\tilde\phi_i\in\mathbb{C}^d\;.
\eeq The expectation value of the witness operator $K_n$ in
$(d\times d)^n$ reduces to an expectation value of $d\times d$
operators \beq \langle
K_n\rangle
&=&\frac{1}{d^2} \frac{1}{d^n}\sum_{k,l} \kappa_{k,l} \sum_{g,h}
\langle\eta_1|
\mathrm{W}_{g,h}|\phi_1\rangle\langle\phi_1|\mathrm{W}_{g,h}^\dagger|\eta_1\rangle\cdot\;\dots\;\cdot
\langle\eta_{n-1}|
\mathrm{W}_{g,h}|\phi_{n-1}\rangle\langle\phi_{n-1}|\mathrm{W}_{g,h}^\dagger|\eta_{n-1}\rangle\\
&&\qquad\qquad\qquad\qquad\cdot\langle\eta_n| \mathrm{W}_{k,l}
\mathrm{W}_{g,h}|\phi_n\rangle\langle\phi_n|\mathrm{W}_{g,h}^\dagger\mathrm{W}_{k,l}^\dagger|\eta_n\rangle
\ist\frac{d^2}{d^2}\cdot C\cdot \;\frac{1}{d^n}\sum_{k,l}
\kappa_{kl} \langle \eta_n|
\mathrm{W}_{k,l}|\tilde\phi_n\rangle\langle\tilde\phi_n|\mathrm{W}_{k,l}^\dagger|\eta_n\rangle\nonumber\eeq
\end{widetext}
with $C \geq 0$. Therefore, $K_n$ is an entanglement witness if the
operator $M_{\phi}=\sum_{k,l}
\kappa_{k,l}\mathrm{W}_{k,l}|\phi\rangle\langle\phi|\mathrm{W}_{k,l}^\dagger$
is not negative for all $\phi\in\mathbb{C}^d$ and it is optimal if
$\det{M_{\phi}}=0$.
\end{proof}


\textbf{Example showing the geometry of separability and PPT--bound
entanglement for different dimensions:} Let us consider any two
vertex states mixed with the totally mixed state, i.e.
\beq\label{exampledim} \rho=\frac{1-\alpha-\beta}{d^2}
\id_d^{\otimes 2n}+ \alpha\; \rho^{\textrm{vertex}\;\otimes
n}_{0,0}+\beta\; \rho^{\textrm{vertex}\;\otimes n}_{0,1}\;.\eeq The
positivity condition of the density matrix on the parameters
$\alpha,\beta$ give three lines which form a triangle. Likewise we
obtain the parameter region for the states which are PPT entangled.
This is visualized in Fig.~\ref{linestates} for dimension $d=2,3,4$.
The authors of Ref.~\cite{BHN1} found by optimizing the witness
operator for bipartite qutrits PPT--bound entanglement if either
$\alpha$ or $\beta$ is negative. By Theorem $2$ this means that we
found a whole region of PPT--bound entanglement for any number of
qutrit pairs $n$.

\begin{figure}
\begin{center}
\includegraphics[width=5cm, keepaspectratio=true]{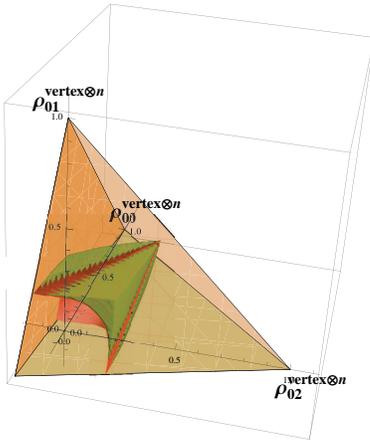}
\caption{Three dimensional slice through the eight dimensional
simplex for $d=3$, given by Eq.~(\ref{example3dim}). The
(transparent yellow) tetrahedron is given by the positivity
condition, the (red) cone represents the PPT condition. Inside the
PPT cone there are also bound entangled states, which cannot be
distilled at all. The (green) Christmas tree shaped area is obtained
via application of the distillation protocol and shows states that
can not be distilled to one of the edge states. This area was
obtained numerically, yet it is most intriguing that it, up to
numerical precision, coincides with the states not detected by
entanglement measures derived from the $m$--concurrence
\cite{HHK1}.}\label{3Dpicture}
\end{center}
\end{figure}

\textbf{Distilling bound entanglement:} While the basic
geometric structure of separable and PPT--bound entangled and
entangled states remains unchanged with $n$, the properties of the
states in the simplex change drastically. They are bound
entangled as the vertices states cannot be distilled (theorem $1$).
However, as we prove in the following for $d=3$ some states inside
$\W^{\otimes n}$ can be distilled by a certain protocol to the
vertices states. Let's consider the following distillation protocol:
\begin{enumerate}
    \item Take a copy of the state: $\rho^{\otimes 2}$, the first dit will be regarded as source, the second as target dit.
    \item Apply the unitary gate $U_m$ in all subsystems: $\rho_T=U_m^{\otimes n}\rho^{\otimes 2}U_m^{\otimes
    n}$ with $U_m:=(1-\delta_{ij})\;|ij\rangle\langle
ij|+\delta_{ij}\;(|ij\rangle\langle im|+|im\rangle\langle ij|)\;.$
    \item Project onto $|m\rangle\langle m|$ in all target systems: $\id_d\otimes|m\rangle\langle m|\;\rho_T\;\id_d\otimes|m\rangle\langle m|$
    \item Discard target dits.
\end{enumerate}
With this protocol it is possible to ``distill'' many NPT-entangled
states in the simplex into a vertex state. Consider e.g. the
following state \beq\label{example3dim}
\rho\ist\frac{1-\alpha-\beta-\gamma}{9} \id_3^{\otimes 2n}+ \alpha\;
\rho^{\textrm{vertex}\;\otimes n}_{0,0}\nonumber\\+\beta\;
\rho^{\textrm{vertex}\;\otimes n}_{0,1}+\gamma\;
\rho^{\textrm{vertex}\;\otimes n}_{0,2}\;.\eeq This is an example of
a so called ``line'' state, where the same Weyl operator connects
all vertex states. This is visualized in Fig.~\ref{3Dpicture}.
Surprisingly, the ``distillable'' states are the ones which are
detected by the bounds on the multipartite qudit measure introduced
in Ref.~\cite{HHK1}.

Clearly, for $n=1$ the vertex states are pure and therefore it is a
genuine distillation protocol, however, for $n\geq 2$ the vertex
states are no longer pure, the protocol distills up to a certain
degree of entanglement and purity. Note, that for $d=2$ and $n=2$
this has already proven to be very useful, as the vertex states can
be used to reduce communication complexity and for remote
information concentration for $2n$
parties~\cite{HorodeckigeneralizedSmolin2}.

%

\textbf{Conclusion:} We have introduced a whole new class of bound
entangled states for arbitrary $n$ pairs of qudits ($d$ degrees of
freedom), the extended simplex $\W^{\otimes n}$, and proven that all
states are non--distillable. The very nature of their bound
entanglement stems from the multipartite construction and may be
unlocked if two parties work together. Inside the simplex ($d\geq
3$) there also exist states which cannot be distilled, because they
are nonseparable PPT--states. Thus in the multipartite and
multidimensional scenario there exist at least two classes of bound
entangled states: those which may be unlocked via multipartite
cooperation and those which cannot be distilled even if two or more
parties cooperate. One could also say the PPT--bound states for any
$n\geq 2$ are \textit{bound--bound} entangled, i.e. PPT--bound and
multi-particle--bound. Moreover, this feature is given for arbitrary
dimensions $d$. In Fig.~\ref{linestates} we showed how the geometry
of separability, PPT and entanglement changes with increasing
dimension $d$. Last but not least our distillation protocol for
$d=3$ shows that almost all $NPT$--entangled states are two copy
distillable to the vertex states and, consequently, the states are
noise resistant. All these special features of these state spaces
may help
to develop novel applications and novel schemes for multipartite quantum communication. 

\end{document}